\documentclass{article}
 \usepackage{amsbsy}
  \usepackage{bm}
 \usepackage{amsfonts}
\usepackage{amssymb}
\usepackage{amsmath}
\usepackage{graphics}
\usepackage{graphicx}
\begin{document}

\title{Hyperons, deconfinement and the speed of sound in neutron stars}
\author{R. M. Aguirre }
\date{\it{Departamento de Fisica, Facultad de Ciencias Exactas}, \\\it{Universidad Nacional de La Plata,} \\
\it{and IFLP, UNLP-CONICET, C.C. 67 (1900) La Plata, Argentina.}}


\maketitle

\begin{abstract}
The effects on the speed of sound in neutron stars due to the
presence of hyperons and a phase transition to deconfined quark
matter is investigated. For this purpose a  composite description
within the covariant field theory is used, it consists of
different models for the hadronic and for the unbound quark
configurations. A phase transition with continuous and monotonous
variation of the equation of state is assumed. The predictions
obtained are contrasted with recent observational data on isolated
neutron stars as well as on binary systems. Only one candidate is
finally obtained from six different descriptions. According to the
present calculations the onset of the hyperons causes the
equilibrium speed of sound to exceed the conformal limit. Based on
recent works on the propagation of g-modes in neutron stars,
different definitions of the speed of sound are considered.

 \noindent
\\

\end{abstract}


\section{Introduction}
The study of the structure and dynamics of compact stars offers
the possibility to test a unified theoretical description covering
the many facets of the strong interaction in combination with
gravitation. Therefore it has been a subject of permanent
interest, but in recent years it has concentrated multiplied
efforts since an important amount of observational data has been
acquired and analyzed.

The possibility to use different experimental techniques to focus
on the same event or the same class of objects has created great
expectations in the specialized community. This is particularly
valid for the study of compact stars, as new and previous
information have created a sketch of different aspects such as the
mass-radius relation, the cooling process, the emission of
gravitational waves from binary mergers, etc. From the theoretical
point of view it is expected that all this input will help to shed
light on some longstanding puzzles and to improve the models and
procedures used.

The evidence of very massive neutron stars  with inertial masses
above 2 $M_\odot$ \cite{DEMOREST,ROMANI,FONSECA}, has introduced
some tension with certain predictions of the relativistic field
theory of hadrons. Calculations made in this framework, using a
mean field approximation, have shown that the emergence of the
hyperon population at densities well above the normal nuclear
density produces an energetically favorable state. The persistence
of the hyperons extends to extremely large densities
and affects significatively the composition of the core of the star.\\
However, most of these results do not admit a neutron star with
mass as high as $M/M_\odot \simeq 2$. This situation is known in
the literature as the hyperon puzzle. \\
A similar picture is obtained when a deconfinement transition is
considered. For this purpose a composite model is usually
employed, corresponding to the hadronic phase and the deconfined
quark-gluon plasma. A first order transition, or even a
coexistence of phases, lowers the energy of the system but also
excludes the minimum upper bound for the star mass in most cases.

Closely related to the determination of the star masses is the
mutual deformation of binary systems due to gravitation, as the
mass distribution of the binary components becomes relevant at
advanced stages of the inspiral process. In particular the
quotient of the quadrupole deformation to the perturbing tidal
field is the only quantity characterizing its influence on the
gravitational wave phase emitted in the early steps
\cite{HINDERER}. A partial answer to this problem was given by the
first detection of a gravitational wave generated by the collapse
of a binary system of neutron stars \cite{ABBOTT1}, from which the
chirp mass of the system was determined with high precision to be
${\cal M}/M_\odot \simeq 1.19$. To obtain information about  the
masses of each component, two different regimes for the spin of
the rotating stars are considered in \cite{ABBOTT1}. Taking the
adimensional parameter $j=c J/G M^2$, where $J$ is the total
angular momentum, they find for $j < 0.89 $ that $1.36 <
m_1/M_\odot < 2.26$ and $0.86 < m_2/M_\odot < 1.36$. Whilst for $j
< 0.05 $ the results are $1.36 < m_1/M_\odot < 1.60$ and $1.16 <
m_2/M_\odot < 1.37$. In addition the tidal deformability for a
neutron star with mass $1.4 M_\odot$ was bounded by
$\Lambda_{1.4}\leq 800 \;(970)$ for the case $j<0.89 \;(0.05)$.
Further refinements \cite{ABBOT2} obtained the preferable values
$1.36 < m_1/M_\odot < 1.62$, $1.15
< m_2/M_\odot < 1.36$, and $\Lambda_{1.4}=190^{+390}_{.120}$.\\

The description of the neutron stars based on microscopic models
of the strong interaction still suffers from important
uncertainties. Intensive work has been devoted to contrast the
recently obtained observational data with the predictions of a
high variety of hadronic models
\cite{TAKAMI,WEIH,NANDI,LOURENCO,SEDRAKIAN,TRAVERSI,GOMES,HAN,DEXHEIMER,LANDRY,HOLZ,ESSICK,FATTOYEV,BLACKER,
LAU,LEGRED,PANG,MOST,FERREIRA,SHANGGUAN,RAAIJMAKERS}. Many of
these studies have  assumed that  matter is composed of protons
and neutrons as the only baryons. Hence the crucial requisite to
accommodate neutron stars with mass at least of $M \simeq 2
M_\odot$ is guaranteed \cite{LOURENCO}. However, in most cases,
the mechanism which inhibits the onset of the hyperons is not
explicitly stated. A smaller number of investigations include
effects of the hyperons
\cite{SEDRAKIAN,TRAVERSI,GOMES,HAN,DEXHEIMER,LANDRY,HOLZ,BLACKER,MOST}.
The possibility of a deconfinement phase transition including
different types of realizations has been the subject of the
thorough study \cite{HAN}, and also of \cite{GOMES,PANG}. For
instance, the outcomes in \cite{HOLZ} about the composition of the
binaries in the GW170817 event depend on the amount of a priori
information deposited on the sampling of equations of state. For
the informed one the probability for a free quark phase is $56\%$
against $44\%$ for a pure hadronic phase. The less informed sample
obtains a reversed $36\%$ against $64\%$, respectively. The
analysis made in \cite{PANG} including data from GW170817 and
GW190425 events, does not find evidence of a strong phase
transition.\\
Transitions including discontinuities in the thermodynamical
potential have received special attention
\cite{WEIH,BLACKER,LAU,MOST} because they present more evident
effects and also they would be detectable by the postmerger
gravitational wave \cite{BLACKER}.

An important feature of the equation of state (EoS) is the
relativistic speed of sound $v_s$ defined by $v_s^2=c^2\,dP/d{\cal
E}$, where $P$ and ${\cal E}$ are the pressure and the energy
density of the system, respectively. The value $v_s/c=1/\sqrt{3}$
has been considered as an upper bound based on very general
arguments. The potential conflict between this assumption and the
existence of massive neutron stars was pointed out in
\cite{BEDAQUE}. This observation has motivated several
investigations on the role of an hypothetical upper limit of the
speed of sound on the properties of neutron stars and binary
systems
\cite{MOUSTAKIDIS,OEVEREN,TEWS,ZHANG,REED,MARGARITIS,KANAKIS}. It
must be noted that \cite{ZHANG} found that some equations of
state, often considered as a paradigm, violates causality when the
baryonic density takes large enough values. Much of these works go
further with the approach used in \cite{ALFORD}, where the EoS is
separated into a low density part described by a favorite model,
and a schematic high density contribution depending on the speed
of sound. The results obtained for the structure of neutron stars
are then contrasted with the observational evidence, mainly the
maximum mass and the tidal
deformation of a pair of interacting stars.\\
The possibility that matter is partially composed of hyperons at
intermediate densities has not been considered in most of these
references. However, \cite{MOTTA} has paid special attention to
this issue. Recently these investigations have been extended by
considering a nontrivial structure for the speed of sound in dense
matter \cite{TAN}, including the effects of hyperons as well
as different types of phase transitions. \\
An interesting consequence of a deconfinement transition in the
scenario of the inspiraling process of a binary system has been
analyzed in \cite{JAIKUMAR}. If the principal mode of the
nonradial gravitational oscillations (g-mode) and the tidal force
are resonant, this could affect the phase of the gravitational
waveform. The frequency of these oscillations depends on the
difference of the squared equilibrium and adiabatic speeds of
sound. If the thermodynamical potentials are continuous through
the transition, but the derivative $dP/d{\cal E}$ shows a finite
discontinuity, the same type of behavior is expected in the
frequency of the g-mode.\\
Thus, the speed of sound plays an important role in the
description of an isolated neutron star as well as for a binary
system. It is a direct measure of the stiffness of the EoS, it
explicitly enters in the definition of the linearized metric
perturbation which is used in the evaluation of the second Love
number, and determines the behavior of the frequency of the g
mode. For these reasons the aim of the present work is to analyze
the speed(s) of sound under several circumstances.

This work is organized as follows: in the next section the general
theoretical description is presented. Section \ref{Sec3} is
devoted to describe the evaluation of some properties of a neutron
star. Specific results are shown and discussed in Sec. \ref{Sec4}
,and finally, the conclusions are drawn in Sec. \ref{Sec5}.
 For the sake of completeness, some calculations
on the quark effective masses are shown in the Appendix.

\section{Theoretical description}\label{Sec2}

To describe the neutron star, several versions of the covariant
field theory of hadrons are used. In these models the baryons
couple linearly to mesons, and the latter exhibit different types
of self-interactions. Thus the Lagrangian density can be written
as
\begin{eqnarray}
{\mathcal L}_H&=&\sum_b \bar{\psi}_b\left(i \not \! \partial -M_b
+g_{\sigma b}\, \sigma+ g_{\xi b} \, \xi+ g_{\delta b}\,
\mathbf{\tau} \cdot \mathbf{\delta} - g_{\omega b} \not \! \omega
- g_{\phi b} \not\! \phi - g_{\rho b} \mathbf{\tau} \cdot \not \!
\mathbf{\rho} \right) \psi_b  \nonumber\\&&+ \frac{1}{2}
(\partial^\mu \sigma
\partial_\mu \sigma - m_\sigma^2 \sigma^2)-\frac{A}{3}\, \sigma^3-\frac{B}{4}
\,\sigma^4 + \frac{1}{2} (\partial^\mu \mathbf{\delta} \cdot
\partial_\mu \mathbf{\delta} - m_\delta^2 \delta^2)+ G_{\sigma \delta} \sigma^2 \,\delta^2\nonumber\\&&
 + \frac{1}{2} (\partial^\mu \xi
\partial_\mu \xi - m_\xi^2 \xi^2)-\frac{1}{4}
 W^{\mu \nu} W_{\mu \nu} + \frac{1}{2} m_\omega^2
\omega^2 +\frac{C}{4} \, \omega^4-\frac{1}{4}
 R^{\mu \nu}\cdot R_{\mu \nu}  \nonumber \\
&&+ \frac{1}{2} m_r^2 \rho^2+ G_{\omega \rho} \rho^2 \, \omega^2
-\frac{1}{4}
 F^{\mu \nu} F_{\mu \nu} + \frac{1}{2} m_\phi^2\,\phi^2\label{LAGRANGIAN}
\end{eqnarray}

\noindent where the sum runs over the octet of baryons. In
addition to the commonly used $\sigma, \omega, \rho$ mesons, here
the scalar iso-vector field $\delta^a$, with $a=1-3$,  as well as
the hidden strangeness $\xi, \phi$ mesons are also included. The
$\delta$ and $\xi$ particles can be identified with the $a_0$
(980) and $f_0$(980) states, respectively. The $\xi, \phi$ are
assumed as mainly composed by a $s \bar{s}$ pair, and therefore,
they couple only to the hyperons. The coupling constants $g_{m b},
m=\sigma, \xi, \delta, \omega, \rho, \phi$ and $ A, B, C,
G_{\sigma \delta}, G_{\sigma \delta}$ vary from one model to
another and are fixed to reproduce different sets of empirical
data. The equations of motion corresponding to this Lagrangian are
solved in the mean field approximation for uniform dense matter,
in a reference frame where the mean value of the spatial component
of the baryon currents are zero. Furthermore, all the degrees of
freedom are considered as stable states of the strong interaction.
Under such conditions the equations are greatly simplified, since
the meson mean values do not vary spatially, and only the third
component of the isomultiplets are nonzero
\begin{equation}
\left(i \not \! \partial -M^*_b - g_{\omega b} \, \omega_0 -
g_{\phi b}\,  \phi_0 - g_{\rho b}\, I_b\, \rho_0 \right)\psi_b=0,
\nonumber \end{equation}
\begin{eqnarray}\left( m_\sigma^2-2\, G_{\sigma \delta}\,\delta^2 \right) \sigma +
A \sigma^2 + B \sigma^3&=& \sum_b g_{\sigma b}\, n_{sb},\nonumber \\
\left( m_\delta^2-2\, G_{\sigma \delta}\,\sigma^2 \right) \delta &=& \sum_b g_{\delta b}\, n_{sb},\nonumber \\
m_\xi^2\, \xi &=& \sum_b g_{\xi b}\, n_{sb},\nonumber
\end{eqnarray}
\begin{eqnarray}\left(m_\omega^2 + C \,\omega_0^2 + 2 \,G_{\omega \rho}\ \rho_0^2\right)
\omega_0
& =& \sum_b g_{\omega b} \,n_b \nonumber\\
\left( m_\rho^2+ 2\, G_{\omega \rho}\, \omega_0^2\right) \rho_0
&=& \sum_b g_{\rho b} I_b \,n_b, \nonumber \\ m_\phi^2 \,\phi_0&=&
\sum_b g_{\phi b} n_b \nonumber
\end{eqnarray}
where $I_b$ is the 3 isospin component, $M^*_b=M_b-g_{\sigma b}\,
\sigma- g_{\xi b} \, \xi- g_{\delta b}\, I_b \delta$ is the
effective mass of the baryon $b$, and the source of the meson
equations are the baryon densities
\[
n_b=  \frac{p_b^3}{3 \pi^2} ,\; \; \; \;n_{sb}= \frac{M^\ast_b}{2
\pi^2} \left[ p_b E_b- M^{\ast 2}_b \ln \left(\frac{p_b+
E_b}{M^\ast_b} \right)\right]\nonumber
 \]
 The left side equation introduces the Fermi momentum, and
 $E_b=\sqrt{p_b^2+M^{\ast 2}_b}$ is used.
 Within the approach, the energy density of the system is given by
\begin{eqnarray}
\mathcal{E}_H&=&\frac{1}{4} \sum_b \left(n_{s b} M^\ast_b + 3 n_b
E_b \right)+\frac{1}{2}\left( m_\sigma^2 \sigma^2 + m_\delta^2
\delta^2 +m_\xi^2 \xi^2 +m_\omega^2 \omega_0^2+ m_r^2 \rho_0^2+
m_\phi^2 \phi^2 \right) \nonumber\\&&+ \frac{A}{3}
\sigma^3+\frac{B}{4} \sigma^4 - G_{\sigma \delta} \sigma^2
\delta^2+ \frac{3}{4} \omega_0^2\left(C \, \omega_0^2 +4 G_{\omega
\rho} \,\rho^2 \right) \nonumber
\end{eqnarray}
The pressure is obtained by the canonical relation \[P=\sum_b
\mu_b\, n_b - \mathcal{E}_H,\] and the chemical potentials are
given by $\mu_b=E_b+ g_{\omega b}\,\omega_0 + g_{\phi b}\,\phi_0+
g_{\rho b}\,I_b\,\rho_0$.

Three specific versions of the general expression
(\ref{LAGRANGIAN}) are used in this work. The first one is based
on the GM1 model of \cite{GLENDENNING} which takes as reference
values $n_0=0.153$ fm$^{-3}$ for the normal nuclear density, and
$E_\text{bind}=-16.3$  MeV, $E_\text{sym}=32.5$ MeV,
$M_N^*/M_N=0.7$, and $K=300$ MeV for the binding energy, the
symmetry energy, the effective nucleon mass, and the nuclear
compressibility in normal conditions. These constraints determine
the constants $g_{\sigma N}, g_{\omega N}, g_{\rho N}, A$ and $B$
which can be consulted in \cite{GLENDENNING}. The model has been
updated in \cite{WEISSENBORN} by introducing the couplings between
hyperon and vector mesons according to the SU(6) symmetry of the
quark model
\[ g_{\omega \Lambda}=g_{\omega \Sigma}=2\,g_{\omega \Xi}=\frac{2}{3} g_{\omega
N}; \; g_{\rho \Lambda}=0, \frac{1}{2}g_{\rho \Sigma}=g_{\rho
\Xi}=g_{\rho N};\]
\[g_{\phi \Lambda}=g_{\phi \Sigma}=\frac{1}{2}\,g_{\phi \Sigma}=-\frac{\sqrt{2}}{3}\,g_{\phi N}.\]
The scalar mesons $\xi, \delta$ are discarded in this scheme, and
all the constants $C, G_{\sigma \delta}, G_{\omega \rho}$ are
taken as zero. The remaining three parameters $g_{\sigma b},\,
b=\Lambda, \Sigma, \Xi$ are determined by adjusting the energy
$U_b=g_{\omega b}\, \omega-g_{\sigma b}\, \sigma$ (subtracted the
vacuum rest mass) of an isolated hyperon at rest, immersed in
isospin symmetric nuclear matter at the normal density . Although
their empirical values are not well known,  they are usually taken
as
\begin{equation} U_\Lambda=-30\, \text{MeV}, \;U_\Sigma=30\, \text{MeV}, \,
U_\Xi=-18 \, \text{MeV}. \label{STANDARD}\end{equation}
 Motivated
by the uncertainty just mentioned, Ref. \cite{WEISSENBORN}
analyzed the effect of the variation of $U_b, b=\Sigma, \Xi$ on
the maximum mass $M_\text {max}$ of an isolated neutron star. It
was found that keeping $U_\Xi$ fixed and varying $-40 < U_\Sigma\,
[$MeV$] < 40$ produce small changes and always $M_\text
{max}<2\,M_\odot$. Inversely if $U_\Sigma$ is kept fixed, it is
found that the maximum mass increases monotonously with $U_\Xi$,
reaching $M_\text {max}=2.04\,M_\odot$ for $U_\Xi=40$ MeV.
Therefore, the more optimistic case with $U_\Xi=40$ MeV is adopted
in this work, obtaining the following values $g_{\sigma
\Lambda}/g_{\sigma N}=0.617$, $g_{\sigma \Sigma}/g_{\sigma
N}=0.404$, and $g_{\sigma \Xi}/g_{\sigma N}=0.113$. The
parametrization thus obtained will be labeled as GM1e.\\
The second model is based on the NL3 parametrization \cite{LALA},
which was adjusted to describe with acceptable accuracy the
properties of several atomic nuclei. It gives the following
results for uniform isospin symmetric nuclear matter $n_0=0.148$
fm$^{-3}$, $E_\text{bind}=-16.3$  MeV, $E_\text{sym}=37.4$ MeV,
$M_N^*/M_N=0.6$, and $K=271.7$ MeV. The original version was taken
further in \cite{YANG}, introducing hyperons in interaction
through the scalar and vector mesons $\xi$ and $\phi$
respectively. As in the previous case the hyperon-vector meson
couplings are related to $g_{\omega N}, g_{\rho N}$ by assuming
SU(6) symmetry and the scalar sector is adjusted to satisfy the
constraints (\ref{STANDARD}). It must be noted that $g_{\xi b}$
does not participate in these relations since they are defined at
zero hyperon density, hence there is some freedom for choosing
these couplings. We adopt here the set labeled as "weak YY" in
\cite{YANG}, so the numerical values of the parameters are taken
from this reference. To complete the comparison with Eq.
(\ref{LAGRANGIAN}) it must be said that the $\delta$ meson is not
considered, hence $G_{\sigma \delta}=0$ and also $C=G_{\omega
\rho}=0$. This prescription will be denoted as NL3e in the
following.

Finally, the third model (M$\sigma \delta$) has recently been
proposed \cite{KUBIS} with the aim of study how the properties of
the neutron stars are affected by mixing interactions between the
$\sigma - \delta$ scalar mesons. The prescription of \cite{MCS} is
adopted to choose the model parameters. Thus the reference
empirical values $n_0=0.16$ fm$^{-3}$, $E_\text{bind}=-16$  MeV,
$E_\text{sym}=32$ MeV, $M_N^*/M_N=0.65$, $K=230$ MeV are taken,
and the slope parameter of the symmetry energy $L=50$ MeV is
reproduced in addition. In the present work this formulation is
complemented with the inclusion of hyperons and the $\phi$ vector
meson. In order to simplify the scheme, the scalar meson $\xi$ is
not considered here and the coupling of the $\delta$ meson to the
hyperons is taken as zero. Eventually the last items will be the
subject of future work. The hyperon-vector mesons are fixed
according the SU(6) symmetry scheme, and the hyperon-scalar meson
couplings follow from (\ref{STANDARD}), obtaining in this way
$g_{\sigma \Lambda}=5.616$, $g_{\sigma \Sigma}=3.989$, and
$g_{\sigma \Xi}=2.920$.

The hypothesis of homogeneous matter which lead to the equations
of motion shown above, is appropriate for a range of densities
above several tenths of the normal nuclear value $n_0$. The
electromagnetic interaction, not included in (\ref{LAGRANGIAN}),
gives rise to nonhomogeneous structures. For this reason the
equation of state evaluated in \cite{BPS} is adopted for the low
density regime and assembled to the results of the interaction
(\ref{LAGRANGIAN}) by imposing continuity at the joining point.\\
At the other extreme, for very dense matter it is expected that
hadrons are no longer the most stable configuration of bound
quarks and a transition to deconfined quarks happens. To take
account of a state of homogeneous quark matter, two different
descriptions are examined in this work, the Nambu Jona-Lasinio
(NJL) and the bag (BM) models. In the schematic BM the
noninteracting quarks have a current mass and a change of scale
due to nonperturbative effects are represented by the bag constant
$B$ added to the thermodynamical potential. The NJL, instead,
presents interacting quarks which generate their own constituent
masses. This effective mass depends on the properties of the
medium and are expected to decrease with increasing
baryonic density. \\
The energy density for both models are  as follows
\begin{equation}
\mathcal{E}_\text{Q}^\text{BM}=\frac{N_c}{\pi^2}\sum_q
\int_0^{p_q} dp\,p^2 \, \sqrt{p^2+m_q^2}+ B, \nonumber
\end{equation}
\begin{equation}
\mathcal{E}_\text{Q}^\text{NJL}=\frac{N_c}{\pi^2}\sum_q \left[
\int_\Lambda^{p_q} dp\,p^2 \, \sqrt{p^2+M_q^2}+ 2 \, G \, n_{s
q}\right]- 4\, K\, n_{s u} \, n_{s d }\, n_{s s}+ \mathcal{E}_0
\label{Enjl}
\end{equation}
where $q=u,\,d,\, s$, $p_q$ is the Fermi momentum which is related
to the baryonic number density by $n_q =p_q^3/3\pi^2$, and $m_q$
is the current quark mass. The total baryon number
density is represented by $n_\text{Q}=\sum_q n_q$. \\
In particular the NJL model uses a cutoff $\Lambda$  for the
momentum integration, $G$ and $K$ are the couplings for four and
six quark interactions, $\mathcal{E}_0$ is a constant introduced
to obtain zero vacuum energy, and the effective masses are given
by
\[M_i=m_i-4\, G\, n_{s i}+2\, K \, n_{s j}\, n_{s k},\; j\neq i \neq k.  \]
The quark condensates $n_{s q}$ can be expressed as
\[ n_{s q}=\frac{N_c}{\pi^2} \,M_q \int_\Lambda^{p_q} \frac{dp \, p^2}{\sqrt{p^2+M_q^2}}\]
The chemical potential for each flavor is simply
$\mu_q=\sqrt{p_q^2+m_q^2}$ within the BM, or
$\mu_q=\sqrt{p_q^2+M_q^2}$ in the NJL.

The transition between the hadronic and deconfined phases has been
described in different dynamical schemes. In this work the picture
of a continuous and monotonous equation of state, with an
intermediate state of coexisting phases is adopted. It is commonly
denominated as the Gibbs construction. If $\chi$ is the spatial
fraction occupied by the deconfined phase, then the total energy
and the baryonic number densities of the system are given by
\begin{eqnarray} \mathcal{E}&=&\chi\,\mathcal{E}_\text{Q} + (1-\chi)\, \mathcal{E}_\text{H},
\\
n&=&\chi\, n_\text{Q} + (1-\chi) \,  n_\text{H}
\label{BDens}\end{eqnarray}

Furthermore, for thermodynamical equilibrium of the coexisting
phases the partial pressures of each phase must coincide
\begin{eqnarray} P_B=\sum_b \mu_b\,n_b - \mathcal{E}_\text{H}=
\sum_q \mu_q\,n_q - \mathcal{E}_\text{Q}\label{PCond}.
\end{eqnarray}

To describe neutron star matter the complementary requirement of
electrical neutrality is imposed. To reach this condition a fluid
of noninteracting leptons (electrons and muons) is considered,
which freely distributes among the hadron and quark phases so that
the condition
\begin{equation}0=3\,\chi \sum_q C_q\,n_q + (1-\chi)
\sum_b C_b \,n_b -\sum_l n_l,\label{ECharge}\end{equation}
is satisfied. In this expression $C_k$ stands for the electric
charge in units of the positron charge.\\
These leptons also contribute to the total energy by
\[\mathcal{E}_\text{L}=\frac{1}{\pi^2}\sum_l \int_0^{p_l} dp\,p^2
\, \sqrt{p^2+m_l^2} \] where $n_l=p_l^3/3 \pi^2$, their chemical
potentials can be written as $\mu_l=\sqrt{p_l^2+m_l^2}$, and the
partial lepton contribution to the pressure is $P_L=\sum_l
\mu_l\,n_l - \mathcal{E}_\text{L}$. Hence, the complete
expressions for the energy and the pressure in the mixed phase are
\begin{eqnarray} \mathcal{E}&=&\chi\,\mathcal{E}_\text{Q} + (1-\chi)\, \mathcal{E}_\text{H}+\mathcal{E}_\text{L},
\label{TotalE}\\
P&=&\mu_\text{B}\, n-\mathcal{E}=P_\text{B}+P_\text{L}
 \label{PRESSURE}.\end{eqnarray}

 The coefficient $\chi$ is obtained by using the conditions of conservation of
 the baryonic number, the electric charge, and thermodynamical
 equilibrium, Eqs. (\ref{BDens}), (\ref{ECharge}) and (\ref{PCond}) respectively. Thus, it is uniquely
 determined for each density of neutral matter in equilibrium at zero
 temperature, and it is a dynamical property of the combination of models
 used.

There are two conserved charges which characterize the global
state of the system, the baryonic number and the electric charge
with associated chemical potentials $\mu_\text{B}$ and
$\mu_\text{C}$ respectively. It must be noted that the last one
does not enter in the intermediate expression of Eq.
(\ref{PRESSURE}) because the total electric charge is zero. Both
chemical potentials can be combined to give the chemical
potentials of all baryons, quarks and leptons circumstantially
present. Therefore they are linearly dependent through the
relations of equilibrium against beta decay.

\section{Properties of the neutron star}\label{Sec3}

The EoS is the main input to determine the structure of a neutron
star. In our approach the relation between $P$ and $\mathcal{E}$
is monotonous and continuous. However its first derivative, i. e.
the speed of sound, presents finite discontinuities at the
threshold of the phase transition. Due to the bijectivity of the
relation $P(\mathcal{E})$ one can write
\[c_s^2=\frac{dP}{d\mathcal{E}}=\frac{dP/dn}{d\mathcal{E}/dn}\]
Using Eqs. (\ref{BDens}), (\ref{PCond}), (\ref{TotalE}) it can be
shown that in the mixed phase is $d\mathcal{E}/dn=\mu_\text{B}$,
so that $dP/dn = n\, d\mu_\text{B}/dn$ and finally
\begin{equation}c_s^2=\frac{n}{\mu_\text{B}}\, \frac{d\mu_\text{B}}{dn}
\label{SSpeed}\end{equation}
Although the expression
(\ref{SSpeed}) is quite simple, the evaluation could be difficult
due to the large number
of degrees of freedom and the constraints imposed.\\
For pure hadronic matter, the derivative in Eq. (\ref{SSpeed})
depends on the number $N$ of baryonic species present in the Fermi
sea,
\begin{equation}\frac{d\mu_\text{B}}{dn}=\frac{\det U}{\det V}\label{DmuDn}\end{equation}
Here $U,\,V$
are square matrices of $N+1$ columns with elements
\[U_{i\, N+1}=\delta_{i,\,N+1};\; V_{i\,
N+1}=1-\delta_{i,\,N+1}\]
\[U_{N+1\, j}=V_{N+1\, j}=(p_j/\pi)^2, 1\leq j\leq
N,\]
\[U_{i\, j}=-V_{i\, j}=H_{i j},\, 1 \leq i, j \leq N\]
where
\begin{eqnarray}H_{i j}&=&\left(\frac{p_j}{\pi}\right)^2\Bigg\{
-\frac{M^*_i}{E_i}\frac{M^*_j}{E_j}\frac{1}{\Delta} \left(\alpha_1
\,g_{\sigma i}\,g_{\sigma j}+\alpha_2 \,g_{\nu i}\,g_{\nu j}-\beta
\,g_{\sigma i}\,g_{\nu j}-\beta \,g_{\sigma j}\,g_{\nu i}\right)
\nonumber
\\&&+\frac{1}{\Delta_V}\Big[(m_\rho^2+2\,G_{\omega \rho}\, \omega_0^2 )\,g_{\omega
i}\,g_{\omega j} +(m_\omega^2+3\,C\,\omega_0^2+2\,G_{\omega
\rho}\, \rho_0^2 )\,g_{\rho i}\,g_{\rho j}\nonumber
\\&&-4\,G_{\omega
\rho}\,\omega\, \rho\,(g_{\omega i}\,g_{\rho j}+g_{\omega
j}\,g_{\rho i})\Big] +\frac{g_{\phi i}\,g_{\phi j}}{m_\phi^2}
+\frac{\pi^2\,C_i \,C_j}{\sum_l \mu_l
\,p_l}\Bigg\}+\frac{p_j}{E_j}\,\delta_{i,j},\nonumber
\end{eqnarray}
\[\alpha_1=m_\sigma^2-2\,G_{\sigma \delta} \delta^2+2\,A\,\sigma+3\,B\,\sigma^2+\sum_b
g_{\sigma b}^2\,\lambda_b, \;\;\alpha_2=m_\nu^2-2\,G_{\sigma
\delta} \sigma^2+\sum_b g_{\nu b}^2\,\lambda_b\]
\[ \beta=\frac{1}{\pi^2}\sum_b g_{\sigma b}\,g_{\nu b}\,\lambda_b-4\,G_{\sigma \delta}\,\sigma\, \delta,\;\;
\lambda_b=\frac{1}{\pi^2}\,\int_0^{p_b}\frac{dp\,p^4}{\left(p^2+M^{*\,
2}_b\right)^{3/2}},\]
\[\Delta_V=(m_\omega^2+2\,G_{\omega\rho}\,\rho_0^2+3\,C\,\omega_0^2)(m_\rho^2+2\,G_{\omega
\rho}\,\omega_0^2)-(4\,G_{\omega\rho}\,\omega_0\,\rho_0)^2,\] and
$\Delta=\alpha_1\,\alpha_2-\beta^2$.\\
The scalar sector of the last expressions must be adapted to the
model used, for the GM1e and NL3e cases one must take $\nu=\xi$,
and $\nu=\delta$ for the M$\sigma \delta$ one.

The structure of Eq. (\ref{DmuDn}) also holds in the coexistence
phase, but in this case is
\[U_{i\, N+1}=S\,\delta_{i,\,N+1},\; 1\leq 1 \leq N+1,\]
\[V_{N+1\,N+1}=\frac{1}{S}\left\{\chi\,N_c\sum_q
\left[\frac{S+3\,C_q\,\left(n_\text{H}-n_\text{Q}\right)}{3\,\pi}\right]^2+
\left(\frac{n_\text{H}-n_\text{Q}}{\pi}\right)^2\,\sum_l \mu_l\,
p_l\right\}, \]
\[V_{i\, N+1}=C_i \frac{n_\text{Q}-n_\text{H}}{S}-1,\,1\leq i \leq N,\]
\[U_{N+1\, j}=V_{N+1\, j}=(1-\chi)\left[S+C_j\,\left(n_\text{H}-n_\text{Q}\right)\right]\,
\left(\frac{p_j}{\pi}\right)^2,\; 1\leq j\leq N,\]
\[U_{i\, j}=V_{i\, j}=H_{i j}-\frac{C_i \,C_j\, p_j^2}{\sum_l \mu_l\,p_l},\; 1 \leq i, j \leq N.\] Where $S=\sum_q n_q\,C_q-\sum_b
n_b\,C_b$ has been used.

Finally, in pure quark matter it is found
\[\frac{d\mu_\text{B}^0}{dn}=3 \pi^2 \frac{3\,\sum_q \mu_q\,p_q\, C_q^2+\sum_l \mu_l\,p_l}
{N_c\,\mu_u\,\mu_d\, p_u\,\left(p_d+p_s\right)+\sum_q \mu_q
p_q\,\sum_l \mu_l p_l}\] within the BM. Whereas in the NJL the
more intricate expression
\begin{eqnarray}\frac{d\mu_\text{B}}{dn}&=&\frac{d\mu_\text{B}^0}{dn}+
M_u'\,\frac{M_u}{\mu_u}+2\,M_d'\,\frac{M_d}{\mu_d}+\frac{3}{\mu_u\mu_d}\nonumber \\
&\times&\frac{p_s \left(M_d\,M_d'-M_s\,M_s'\right) \sum_q p_q
\mu_q\,C_q+ \left(2\,p_u \mu_u+ \sum_l p_l \mu_l\right)
\left(M_u'\,\frac{M_u}{\mu_u}+2\,M_d'\,\frac{M_d}{\mu_d}\right)
\sum_l p_l \mu_l}{N_c\,\mu_u\,\mu_d\,
p_u\,\left(p_d+p_s\right)+\sum_q \mu_q p_q\,\sum_l \mu_l
p_l}\nonumber
\end{eqnarray}
is obtained  due to the variation of the constituent quark masses.
Explicit formulas for these quantities are given in the Appendix.

The adiabatic speed of sound $v_a=\partial \mu_B/
\partial n$ enters in the evaluation of the frequency of nonradial
oscillations in compact stars \cite{JAIKUMAR}. It is obtained by
imposing the condition that the relative population of each
fermion species remains constant, even if $\beta$ equilibrium is
not verified. All the physical
constraints are imposed after evaluation of the derivatives. \\
In pure hadronic matter the following result holds
\begin{eqnarray} v_a^\text{H}=&&\frac{1}{n\,\mu_\text{B}}\Bigg\{\sum_b\frac{p_b^2}{3}\,\frac{n_b}{E_b}
+\sum_l\frac{p_l^2}{3}\,\frac{n_l}{\mu_l}
-\frac{1}{\Delta}\left(\alpha_1\,X_\sigma^2+\alpha_2\,X_\nu^2-2\,\nu\,X_\nu\,X_\sigma\right)
+m_\phi^2\,X_\phi^2 \nonumber\\
&+&\frac{1}{\Delta_V}\left[(m_\rho^2+2\,G_{\omega\rho}\,\omega_0^2)\,X_\omega^2+
(m_\omega^2+3\,C\,\omega_0^2+2\,G_{\omega
\rho}\,\rho_0^2)\,X_\rho^2-8\,G_{\omega\rho}\,X_\omega\, X_\rho
\right]\Bigg\}\nonumber \end{eqnarray} where \[X_\alpha=\sum_b
g_{\alpha b}\,n_b\,\frac{M^*_b}{E_b}\;\, \text{for}\,
\alpha=\sigma,\,\delta,\, \xi;\;\; X_\alpha=\sum_b g_{\alpha
b}\,n_b\; \,\text{for}\, \alpha=\omega,\,\rho,\, \phi.\] The
treatment of pure quark matter gives, instead
\[v_a^\text{Q}=\frac{1}{3 n \mu_\text{B}}\left(\sum_q p_q^2\,\frac{n_q}{\mu_q}
+\sum_lp_l^2\,\frac{n_l}{\mu_l}\right) \]
within the BM, and
\[v_a^\text{Q}=\frac{1}{3 n \mu_\text{B}}\left(\sum_q p_q^2\,\frac{n_q}{\mu_q}
+\sum_lp_l^2\,\frac{n_l}{\mu_l}\right)+\frac{1}{\mu_\text{B}}\sum_q
n_q M_q' \frac{M_q}{\mu_q}\] for the NJL.\\ In the domain of
coexistence of phases one can write
$v_a=\chi\,v_a^\text{Q}+(1-\chi)\,v_a^\text{H}$.

The structure of an isolated neutron star can be solved using the
Tolman-Oppenheimer-Volkov equations for the spherically symmetric
case
\begin{eqnarray}
\frac{dP}{dr}&=&- (G/c^2) \frac{[{\cal E}(r)+P(r)]\,[m(r)+4 \pi
r^3 P(r)/c^2]}{r^2[1-2 (G/c^2) m(r)/r]} \, , \nonumber \\
{\cal M}(r)&=&\int_0^r 4 \pi \, {r'}^2 \, [{\cal E}(r')/c^2] \,
dr' \, . \nonumber
\end{eqnarray}
Starting from given values of the central pressure and energy,
these equations are integrated outward until a radius $R$ is
reached for which  $P(R)=0$, and the total mass is defined as
$M={\cal M}(R)$. \\
Once the mass ${\cal M} (r)$ and pressure $P(r)$ distributions
inside the star have been determined one can evaluate the second
Love number $k_2$. For this purpose the radial function $y(r)$,
related to the tidal field, must be found by solving the
differential equation
\[ y'(r)+y^2(r)+f(r)\,y(r)+q(r)\,r^2=0\]
subject to the condition $y(0)=2$. The following definitions have
been used
\begin{eqnarray}
f(r)&=&\frac{1+4\,\pi\,r^2\left(P-{\cal E}\right)}{1-2\,{\cal
M}/r} \nonumber\\
q(r)&=&\left[4 \pi \left(5\,{\cal E}+9\,P+\frac{P+{\cal
E}}{v_s^2}\right)-\frac{6}{r^2}-\frac{4}{r^4}\frac{\left({\cal
M}+4\,\pi\,r^3 P\right)^2}{1-2\,{\cal
M}/r}\right]/\left(1-2\,{\cal M}/r\right) \nonumber
\end{eqnarray}
Then, the Love number is given by
\begin{eqnarray}k_2&=&\frac{8}{5} x^5 (1-2 x)^2 [2-y_R+2 x (y_r-1)]/\Big\{6 x [2-y_R+x (5 y_R-8)]+4 x^3 [
13-11 y_R+x (3 y_R-2)\nonumber \\&+&2 x^2 (1+y_R)]+3 (1-2
x)^2[2-y_R+2 x (y_R-1)]\ln(1-2 x)\Big\}\nonumber \end{eqnarray}
where $x=M/G R$ and $y_R=y(R)$. The tidal deformability is
obtained in this approach as $\Lambda=2\,k_2/3 x^5$.

\section{Results and discussion}\label{Sec4}

In this section a comparative analysis of the results provided by
the different models is made. In first place I focus on the
evidence that can be found in the speed of sound propagating in a
neutron star, about the emergence of exotic degrees of freedom.
Furthermore an analysis is presented on the ability of the
proposed framework to accommodate  the observational evidence
about compact stars.

Numerical evaluation of the hadronic properties has been done by
using the parameter sets discussed in Sec. 2. For the quark sector
I use either the BM with parameters $B=200$ MeV/fm$^3$,
$m_u=m_d=5$ MeV, $m_s=150$ MeV or the NJL model with the SU(3)
parametrization given in \cite{HATSUDA}.

Since there are many works that disregard the role of the hyperons
in high density matter and with the purpose of contrast with this
standard approach, for each of the hadronic models previously
described I consider the case with hyperons
artificially suppressed (NH).\\
 The EoS obtained in different schemes is shown in
Fig.1. In each panel a low energy regime can be distinguished
where all the curves coalesce. It is composed by pure nuclear
matter and leptons. In the high energy extreme, instead, two
different curves indicate the emergence of the pure quark phase as
described by the BM or the NJL models. In general the BM reaches
this instance with lower pressures but steeper slope, i.e. with
higher speed of sound. Between these extremes and as the energy
increases, the emergence of the $\Lambda$ hyperon and of
deconfined quarks  takes place, in the mentioned order. For the
NL3e and M$\sigma \delta$ the heavier $\Xi^-$ is also present, but
in every case no new hyperon species appear during the
coexistence. On the contrary, in the M$\sigma \delta$ description
the preexistent $\Xi^-$ population
extinguishes before the phase transition is completed.\\
It is evident that for all the hadronic models the combination
with the BM produces an EoS softer than that corresponding to the
NJL. The only exception is found for pure quark matter at
extremely high energies, corresponding to densities above $2.5
\times 10^{15}$ g/cm$^3$.\\
 Comparing results with or without hyperons a common pattern is
found, for lower values of  ${\cal E}$ the pressure is slightly
higher for the NH case, but beyond a particular value the relation
is inverted. The point where this change happens is located in the
coexistence phase. The enhancement of the pressure for relatively
high energy in the case of matter containing hyperons, is
particularly noticeable  when the NJL model is used. This feature
deserves special emphasis because it corresponds to a regime
accessible to the core of massive neutron stars.\\
A contrast of the different models shows that in the NL3e the
deconfinement starts at a lower pressure and has a stronger
softening effect on the EoS. Furthermore the coexistence region is
wider for the BM than for the NJL model, thus pure quark matter
appears earlier in the last case. Using this model I have verified
that the threshold for the transition increases with the value
chosen for the parameter $B$ of the bag model. Thus in order to
obtain stable pure hadronic matter for densities below $5\times
10^{14}$ g/cm$^3$ (around twice normal nuclear density) the bound
$B \geq 200$ MeV/fm$^3$ must be satisfied. To give a uniform
treatment, the same value $B = 200$ MeV/fm$^3$ is used in
combination with all the hadronic descriptions.\\
In regard to the remaining models, the critical pressure is lower
for the GM1e than for the M$\sigma \delta$, and the slope of the
EoS is higher in the last case. Therefore it is expected that the
transition to the outer crust of the neutron star develops more
rapidly in the M$\sigma \delta$ model.\\
The pressure at the density $n/n_0=2$ has been estimated in
\cite{ABBOT2} as $P=3.5^{+2.7}_{-1.7}$ dyn/cm$^2$ in order to be
consistent with the observational data obtained in the GW170817
event. In addition several constraints have been imposed to the
EoS, as for instance causality, thermodynamic stability of the
star, same EoS for both components of the binary system, and
consistency with a maximum mass $M_\text{max}/M_\odot=1.97$. For
the calculations in this work the pressure in the NL3e model
exceeds the upper limit by a small $3 \%$ when hyperons are
present and by more than $7 \%$ in the NH case. The remaining
models do not present hyperons  at such density, and the pressure
is greater than the reference value $3.5$ dyn/cm$^2$, by a $20 \%$
in the M$\sigma \delta$ and by a $40 \%$ in the GM1e case.
Notwithstanding the predicted values are consistent with the
experimental bounds.

 The different sets of equations of state are used in the following to study the structure of a
nonrotating neutron star, for a reasonable range of central
pressures. The relation $M(R)$ is shown in Fig. 2 for all the
models considered here. An immediate conclusion is that the use of
the BM always gives smaller masses, and the predicted maximum mass
is far from the empirical bound $M/M_\odot \simeq 2$. The
combination with the NJL, instead, produces admissible results. In
such case the NH approach systematically obtains greater maximum
mass corresponding to a greater star radius. In Table I the
numerical outcomes are summarized. In the following
only the models predicting $M_\text{max}/M_\odot \geq 1.98$ are considered.\\
For all the approaches shown in this table the neutron star with
the standard $M/M_\odot \simeq 1.4$ mass is totally composed of
nucleons. The only exception corresponds to the prediction of the
M$\sigma \delta$ model which finds a tiny $3 \%$ of $\Lambda$
hyperons in the core of the star. In contrast, the central region
of the star with maximum mass is in the coexistence phase. There
are strange degrees of freedom present in the form of hyperons or
as deconfined quarks. \\
Regarding the star with $M/M_\odot = 1.4$, the results for its
radius can be compared with the estimates provided by
\cite{RAAIJMAKERS}. Both, GM1e and NL3e predictions are well above
the upper bound established in that work. The M$\sigma \delta$,
instead, gives $R=12.6$ km that is compatible with the ranges
$12.33^{+0.76}_{-0.81}$ km and
$12.18^{+0.56}_{-0.79}$ km obtained by different approaches in \cite{RAAIJMAKERS}.\\
For further comparison one can take  the Bayesian analysis
presented in \cite{RILEY} for the massive pulsar PSR J0740+6620,
which obtains the radius $R=12.39^{+1.30}_{-0.98}$ for the star
with $M/M_\odot=2.072^{+0.067}_{-0.066}$. Focusing on the M$\sigma
\delta$ model, it gives a star with maximum mass $M/M_\odot=2.003$
with a radius $R=12.097$ km, while the NH case indicates that for
a star with $M/M_\odot=2.07$ corresponds a radius $R=12.67$ km.
Although the mass of the first result is slightly below the
interval given by \cite{RILEY}, the radius of both instances are
in well agreement with the range expected by that work. The
reliability of the values obtained with the M$\sigma \delta$ is
also confirmed by contrasting with \cite{MILLER}. In that work the
information on PSR J0740+6620 is extended by including additional
empirical data to estimate the range $R=12.45 \pm 0.65$ km for
$M/M_\odot=1.4$ and $R=12.35 \pm 0.75$ km for the PSR J0740+6620.

In Fig. 3 the rich structure of the speed of sound for the five
models selected is shown in terms of the baryonic density. The
panel (a), corresponding to the GM1e NH in combination with the
NJL, makes evident some common features, $v_a$ is continuous but
$v_s$ presents finite discontinuities at the borders of the
coexistence domain, as it has already been noted in
\cite{JAIKUMAR}. Furthermore, both definitions are almost
coincident for low densities and seems to converge to a common
value for extremely large densities. In the regime of pure quark
matter the variation of $v_s$ is around $6 \%$ of the speed of
light for the range of densities shown in that figure. In the
center of the maximum mass star, corresponding to $n/n_0 \simeq
4.9$, a change $\Delta{\cal E} \simeq 1$ fm$^{-4}$ causes a drop
in $v_s$ of around $8 \%$ the speed of light. The panels (b) and
(c) additionally incorporate the effect of hyperons. In such case
the general trend of $v_a$ is not highly modified, but $v_s$
reflects markedly the onset of the hyperons that happens before
the deconfinement transition. The $\Lambda$ baryon appears at
$n/n_0=1.9 \; (2.3)$,  while the heavier $\Xi^-$ starts at
$n/n_0=2.2 \; (2.8)$ in the NL3e (M$\sigma \delta$) model.  The
corresponding curve shows a peak followed by a pronounced drop
associated with the jumping-off point of each hyperon. The same
kind of structure has been observed in \cite{MOTTA} at the rise of
the hyperon population, using a model of composite baryons which
does not take account of the deconfinement process.
Significatively the peak values are almost coincident
$v_\text{max}/c=0.63,\,0.61$ and $0.65$ in \cite{MOTTA}, for the
NL3e and M$\sigma \delta$ models, respectively.
\\
Beyond these characteristic configurations, the speed of sound
increases until the beginning of the phase transition, where a new
 drop takes place through a discontinuous jump.  The sudden decrease after these peaks
 induces a noticeable splitting of $v_a$ and $v_s$.
\\
A comparison with the conformal limit $v_\text{lim}=c/\sqrt{3}$,
shows that it is exceeded at the onset of the coexistence of
phases in the models GM1e NH and M$\sigma \delta$ case. In the
latter case the difference becomes noticeable in the NH approach
where the increment $\Delta v_s \simeq v_\text{lim}/3$ is reached
at the threshold density $n_t$. When hyperons are included in the
same framework, $n_t$ is shifted to higher values and the relative
difference is considerably reduced. However, additional points
appear where $v_s > v_\text{lim}$, corresponding to the onset of
the $\Lambda$ and $\Xi^-$ hyperons. In the first case the excess
is as important as in the deconfinement point. The maximum values
of $v_s$ obtained in our calculations and discussed in the
preceding paragraph must be completed with the cases GM1e NH, NL3e
NH and M$\sigma \delta$ NH giving $v_\text{max}/c=0.62,\,0.64,\,
0.78$ respectively. It must be pointed out that both instances of
the M$\sigma \delta$ calculations verify $v_\text{max}/c\geq
0.63$, a value that has been proposed as a minimum upper bound for
the speed of sound
\cite{ALSING}.\\
These observations seem to corroborate the relation between the
magnitude of the speed of sound and the number $N$ of effective
degrees of freedom. In agreement with the general belief, an
increase of $N$ with the density is locally reflected by a sudden
drop in $v_s$, which is realized through a finite discontinuity in
the case of the phase transition. The growth of $v_s$ observed
between these particular points is consistent with the
monotonously increasing trend found in \cite{MOUSTAKIDIS}, where a
variety of nuclear matter equations of state are analyzed.  \\
 The difference $\tau=1/c_s^2-1/c_a^2$ directly affects  the
frequency of the nonradial oscillations of a compact star known as
gravity modes \cite{JAIKUMAR}. For this reason Fig. 4 is devoted
to show $\tau$ as a function of the baryonic density for the
M$\sigma \delta$ model including or not the hyperons. In the upper
panel a detail of the numerator $c_a^2-c_s^2$ is presented, where
the case NH is suitable for comparison with Fig. 6 of
\cite{JAIKUMAR}. The monotonous decrease in the coexistence zone
and the higher values corresponding to the low density threshold
in the present calculations contrast with the results shown in
that figure. The inclusion of hyperons has the notorious effect of
a sudden rise preceding the discontinuity at $n_t$. In the lower
panel the full factor $\tau$ is presented. The discontinuities,
and a preceding staircase structure in the full hyperon treatment,
stand out for medium densities. They are diminished by the strong
drop experienced at the end of the coexistence region. However
this regime would not be reached since according to the present
calculations the pure quark phase state is not realized even for
the most massive neutron star.

As a final item, I analyze the predictions for the tidal
deformability on the members of a binary system. Taking as a
reference a neutron star with $M/M_\odot= 1.4$ the results
obtained for the tidal deformability are shown in the last column
of Table I. Both, GM1e and NL3e results are far beyond the bound
$\Lambda_{1.4}=190^{+390}_{-120}$, suggested in \cite{ABBOT2}. The
results of both instances of the M$\sigma \delta$, instead, are
admissible according to the same criterium but are close to the
upper limit. For all the cases considered the central density
$n_c$ of the reference star is relatively small $1.9 \leq n/n_0
\leq 2.5$, and the conventional degrees of freedom of nuclear
physics are the main ingredients of this low mass star. Therefore
the result for $\Lambda_{1.4}$ is intrinsic to the parametrization
of the models, since the range of densities are reasonably close
to the reference point $n/n_0=1$. It is surprising that the model
inspired in the NL3 parametrization, which accurately describes
the structure of several atomic nuclei, has the worst disagreement
with the empirical expectations.\\
Another parameter of interest is the combined tidal deformability
\[\tilde{\Lambda}=\frac{16}{13}\,\frac{\Lambda_1\,(M_1+12\,M_2) M_1^4+
\Lambda_2\,(M_2+12\,M_1) M_2^4}{(M_1+M_2)^5}\] where $M_i,
\,\Lambda_i$ are the mass and the tidal deformability of the
individual components. On the other hand the chirp mass, given by
the relation
\[{\cal M}^5=\frac{M_1^3 M_2^3}{M_1+M_2},\]
has been determined with accuracy \cite{ABBOT2} for the event
GW170817, while the possible values for $M_1$ are expected to
range within $1.3 < M_1/M_\odot< 1.6$, assuming  $M_2<M_1$. Under
this constraint I have evaluated $\tilde{\Lambda}$ in terms of
$M_1$ for the combined M$\sigma \delta$ and NJL models. The
result, as shown in Fig. 5, lies between $566 < \tilde{\Lambda} <
608$, which is compatible with the expectations for the low spin
prior $\tilde{\Lambda} \in (70,800)$ as well for the high spin
prior $\tilde{\Lambda} \in (0,630)$ \cite{ABBOT3}.\\
In the present calculations a coexisting phase of confined and
unbound quarks is assumed, which can be interpreted as a
consequence of a vanishing interface tension $\sigma_T$. At the
opposite extreme, for very large $\sigma_T$, a discontinuous
transition takes place according to the Maxwell construction. For
intermediate values a nonhomogeneous phase is expected, which can
affect the neutron star properties.  These effects have been
analyzed in \cite{XIA} within a specific model, concluding that
all of them, the maximum mass, the radius, and the combined tidal
deformability monotonously increase with $\sigma_T$. An estimation
of the maximum variation due to finite tension is given there as
$\Delta M_\text{max}/M_\odot=0.02,\, \Delta R=0.6$ km, and $\Delta
\tilde{\Lambda}/\tilde{\Lambda}=0.5$ \cite{XIA}. Thus a scarce
increase in the maximum mass can be obtained at the cost of a
small growth of the radius and a considerable increment of the
tidal deformability.

The present calculations indicate that the properties of the
standard $M/M_\odot=1.4$ star are determined exclusively by the
hadronic EoS, while the deconfined quark EoS could affect the
structure of the more massive stars. Therefore the effect of new
configurations in the deconfined phase, such as superconductivity,
are of interest for determining the upper limit of the neutron
star masses. A large number of studies have focused the effects of
superconducting quark matter on the properties of compact stars
\cite{ALFORD2,BUBALLA,LAWLEY,PAGLIARA,PAULUCCI}. For instance in
\cite{ALFORD2} an effective nuclear model is used in combination
with a BM including a color-flavor locked superconducting phase.
For the latter model the parameters are taken as $B=137$
MeV/fm$^3$, $m_s=200$ MeV and $\Delta=100$ MeV for the energy gap.
The mass-radius relation for the neutron star shows the
significant fact that a sharp quark-hadron phase transition leads
to an unstable star structure. In contrast, the continuous phase
transition allows the existence of stable configurations with
unbound quarks. In any case the maximum mass is slightly reduced
as compared with the unpaired case. This behavior is qualitatively
confirmed in \cite{BUBALLA} where the dynamics of the deconfined
quarks is determined by the NJL within two different
parametrizations. Since the quark-quark interaction is unknown,
the authors assume the same coupling constant as in the four
fields quark-antiquark interaction $G_D=G$. They only consider a
sharp hadron-quark phase transition and also include the
possibility of a light quark superconducting phase (2SC), with
unpaired strange flavor, in addition to the just mentioned
color-flavor locked arrangement. In this case, the presence of the
intermediate two-flavor pairing introduces a narrow window of
stability before the color-flavor locked phase becomes preferable.\\
These types of instabilities have been related to the lack of
confinement of the NJL model \cite{BALDO}, and attributed to the
value of the constant $\mathcal{E}_0$, see Eq. (\ref{Enjl}), used
to render zero the energy density at zero baryonic density. This
argument has been examined in \cite{PAGLIARA}, where a different
procedure to fix the additive constant  has been proposed. With
this modified constant $\mathcal{E}_0^*$, an intermediate stable
2SC phase was found, as in \cite{BUBALLA}. A further increase of
the pairing coupling constant to $G_D=1.2 \,G$, in combination
with $\mathcal{E}_0^*$, extends the range of stability to embrace
the color-flavor locked phase. At the same time the allowed
maximum mass for neutron stars is reduced \cite{PAGLIARA}. \\Based
on these results one can conclude that the inclusion of a
superconducting quark phase, if stable, will lead to a decrease of
$M_\text{ max}$.

\section{Summary and Conclusions}\label{Sec5}

This work is devoted to the study of dense matter at zero
temperature, as can be found in the interior of neutron stars. To
describe the low and medium densities regime, three models of the
field theory of hadrons are used. They have different motivations,
while GM1 and the recent M$\sigma \delta$ focus on bulk properties
of homogeneous matter, the NL3 was calibrated to study atomic
nuclei. In all the cases the formulation has been extended to
include hyperons. For high densities a scheme of deconfined quarks
are considered using either the Bag or the NJL models. In the
first case the quarks do not interact and vacuum effects are
explicitly included through the bag constant $B$, while in the
well known NJL there is a strong interaction between quarks which
give them their constituent masses. In between a coexistence of
phases is assumed which allows a continuous variation of the
thermodynamic potential. As an alternative the situation (NH) with
hyperons artificially suppressed is also taken into account. In
this context the equation of state has been analyzed, and the
speed of sound in particular. The effects on the structure of a
neutron star have been emphasized and the contrast with recent
observational data has been done. \\
For all the cases considered the hyperons emerge near $n/n_0=2$,
and the deconfinement transition starts at a higher density $n_d$
which depends on the model used (see Table I). The well known fact
that the NH approach gives the harder EoS has been corroborated
for each model. Furthermore the combination with the BM
systematically gives a softer EoS as
compared with the NJL case.\\
The adiabatic speed of sound $v_a$ shows a continuous behavior,
while the equilibrium velocity $v_s$ presents finite
discontinuities at the extreme points of the coexistence region
\cite{JAIKUMAR}. I have found that the onset of the hyperons also
has noticeable effects, giving place to characteristic breaks of
the monotonous variation of $v_s$. The quantity $1/v_s^2-1/v_a^2$
which enters in the construction of the frequency of g-mode
oscillations of a star, also has distinctive behaviors according
to the presence, or not, of the hyperons.\\
The relation mass-radius of an isolated neutron star has been
examined and I find that the combinations with the BM are not able
to satisfy the requisite $M_\text{max}/M_\odot > 1.95$, hence
these instances are discarded.\\ Focusing on a star with the
canonical mass $M/M_\odot=1.4$, it is found that its central
density is small enough such that only nucleons and leptons are
present in its composition. The exception is the M$\sigma \delta$,
which predicts a scarce amount of $\Lambda$ hyperons. When
considered as a part of a binary system, its tidal deformability
is expected to be bounded by $\Lambda_{1/4} < 580$ \cite{ABBOT2},
however only the prediction of the M$\sigma \delta$ model
$\Lambda_{1/4}=527$, adjust this condition. Furthermore, when the
experimental value for the chirp mass ${\cal M}/M_\odot=1.186$ is
taken into account, the composed tidal deformability has been
found to satisfy $566 < \tilde{\Lambda} < 608$ which is compatible
with the observational
evidence \cite{ABBOT3}.\\
One can conclude that the formulation of a hadronic model which
includes the hyperons and the realization of a coexistence phase
with deconfined quarks, is absolutely compatible with the recent
experimental data on compact stars. In the case analyzed in this
work, the model denoted as M$\sigma \delta$ \cite{KUBIS,MCS},
achieves this purpose with simplicity by introducing only one
additional term to those commonly used in the field theory of
hadrons. This term consists of a nonlinear meson vertex with a
constant coupling, and continues the long-standing strategy of
representing high density effects by this type of interactions
\cite{BOGUTA}. There are still several open questions about this
model which deserve investigation, as for instance, the
combination of the $\delta$ and the hidden strangeness $f_0(980)$
mesons as mediators of the hyperon interaction, or the
compatibility with the phenomenology of atomic nuclei.

\section*{Acknowledgements}This work was partially supported by the
CONICET, Argentina under grant PIP-616.

\section{Appendix: Mass derivatives in the NJL model}

The derivatives of the constituent quark masses in the NJL are
given by
\[M'_i=-4\,G\,n'_{s i}+ 2 \,K\, \left(n_{s j}\,n_{s
k}\right)',  \;\text{where}\, i\neq j, \,i\neq k,\, j\neq k .\] In
turn the derivatives of the quark condensates are the solutions of
a linear set of algebraic equations
\[ \sum_j {\cal A}_{i j}\,n'_{s j}= \frac{p_i}{D}\,\frac{M_i}{\mu_i}\,{\cal N}_i,\,i=u,d,s\]
where $D=a_u\,\left(p_d+p_s\right)-\left(a_d+a_s\right)\,p_u$,
${\cal N}_u=-\left(a_d+a_s\right)\,\pi^2,\,{\cal N}_d={\cal
N}_s=a_u\,\pi^2$,
\[a_u=\frac{1}{\mu_u}+\frac{2 p_u}{\mu_e \sum_l p_l}, \;a_d=-\frac{1}{\mu_d}-\frac{p_d}{\mu_e \sum_l
p_l}, \; a_s=-\frac{p_s}{\mu_e \sum_l p_l}\] and
\[{\cal A}_{i j}=\left(\frac{\pi^2}{3}-4 G F_i \right)\,\delta_{i j}+
2 K F_i n_{s m} \left(1-\delta_{i j}\right)+2 \,\frac{p_i}{D}
\frac{M_i}{\mu_i}\left(2 G {\cal P}_{i j} \frac{M_j}{\mu_j}-K
\sum_{k\neq j} n_{s l}\,{\cal P}_{i k} \frac{M_k}{\mu_k} \right)\]
where $j \neq m\neq i$ and $k \neq l\neq i$, and
\begin{eqnarray} {\cal P}=\left(
\begin{array}{ccc}
-p_d-p_s& p_d & p_s \\
p_u & -p_u+\left(a_u p_s-a_s p_u\right)\, \mu_d & \left(a_u
p_s-a_s p_u\right)\, \mu_d \\
p_u &-p_u+\left(a_u p_d-a_d p_u\right)\, \mu_d & \left(a_d p_u-a_u
p_d\right)\, \mu_d
\end{array}\right) \nonumber
\end{eqnarray}

\newpage
\begin{table}
\begin{tabular}{l|ccccc}
Model &$n_d/n_0$ &$M_\text{max}/M_\odot$ & $R$ [km]& $R_{1.4}$ [km]& $\Lambda_{1.4}$\\
\hline
GM1e NH Bag &2.36 &1.79 & 12.8 & -- & -- \\
GM1e  Bag &2.40 &1.76 & 12.7 & --  & --\\
GM1e NH NJL &2.55 &1.98 & 13.0 & 13.9 & 910.03 \\
GM1e NJL &2.85 &1.92 & 12.7 & 13.9 & --\\
NL3e NH Bag &1.98 &1.91 & 14.1 & -- & --\\
NL3e Bag &2.04 &1.88 & 14.0 & -- & --\\
NL3e NH NJL & 1.98 &2.08 & 14.3 & 14.8 & 1284.20 \\
NL3e  NJL &3.44 &2.02 & 13.9 & 14.8  & 1284.20\\
M$\sigma \delta$ NH Bag &2.66 &1.90 & 12.5 & -- & -- \\
M$\sigma \delta$ Bag &3.29 &1.86 & 12.2 & --  & --\\
M$\sigma \delta$ NH NJL &3.04 &2.10 & 12.5 & 12.6 & 527.08 \\
M$\sigma \delta$ NJL &4.02 &2.00 & 12.1 & 12.6 & 527.20 \\
\end{tabular}
\caption{\footnotesize The threshold density for the deconfinement
transition, the isolated neutron star properties maximum mass and
its corresponding radius for all the models considered. In the two
last columns the radius and the tidal deformability of a star with
the canonical mass $M/M_\odot=1.4$ for selected cases.}
\end{table}

\newpage
\begin{figure}
\includegraphics[width=0.8\textwidth]{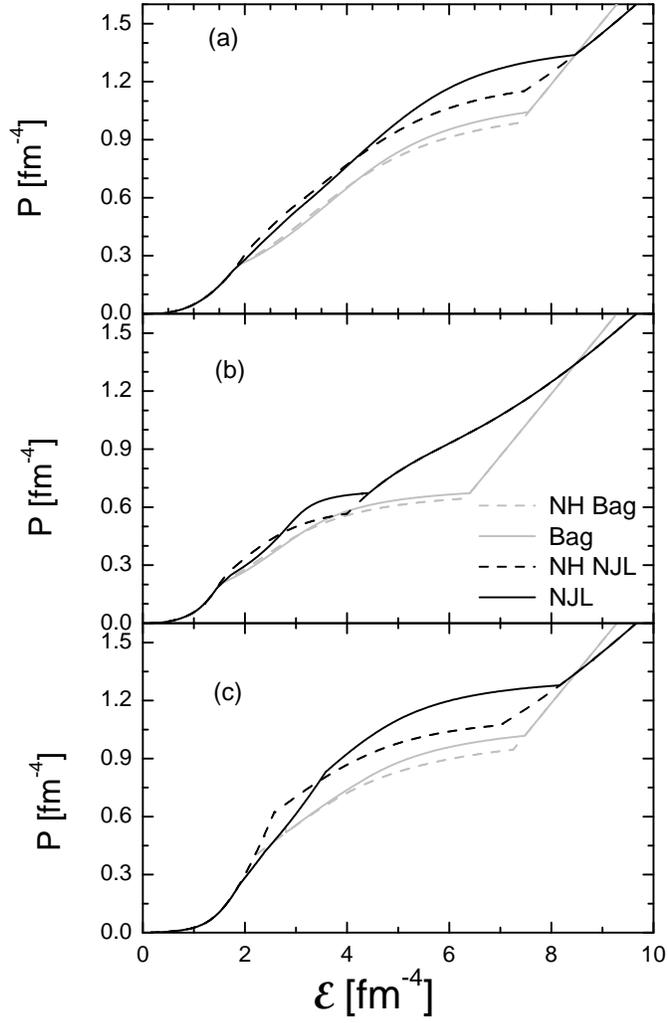}
\caption{\footnotesize The equation of state for the hadronic
models GM1e (a), NL3e (b), and M$\sigma \delta$  (c). The cases
with or without hyperons (NH) have been distinguished according to
the line convention shown. }
\end{figure}

\newpage
\begin{figure}
\includegraphics[width=0.8\textwidth]{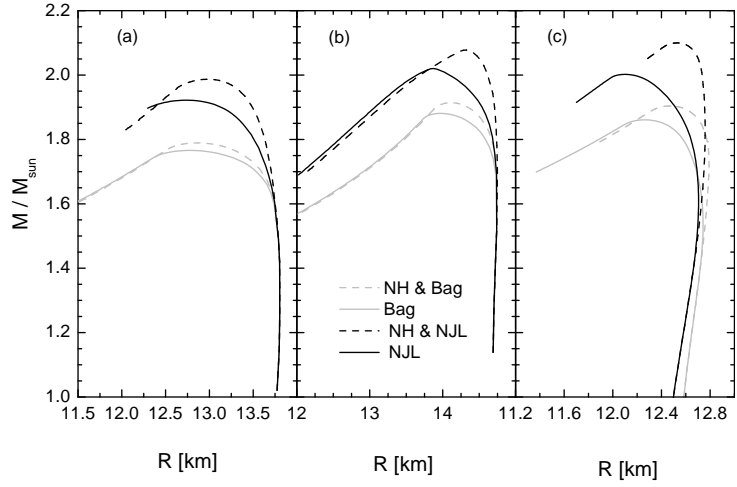}
\caption{\footnotesize The mass-radius relation for an isolated
neutron for the hadronic models GM1e (a), NL3e (b), and M$\sigma
\delta$ (c). The cases with or without hyperons (NH) have been
distinguished according to the line convention shown. }
\end{figure}

\newpage
\begin{figure}
\includegraphics[width=0.8\textwidth]{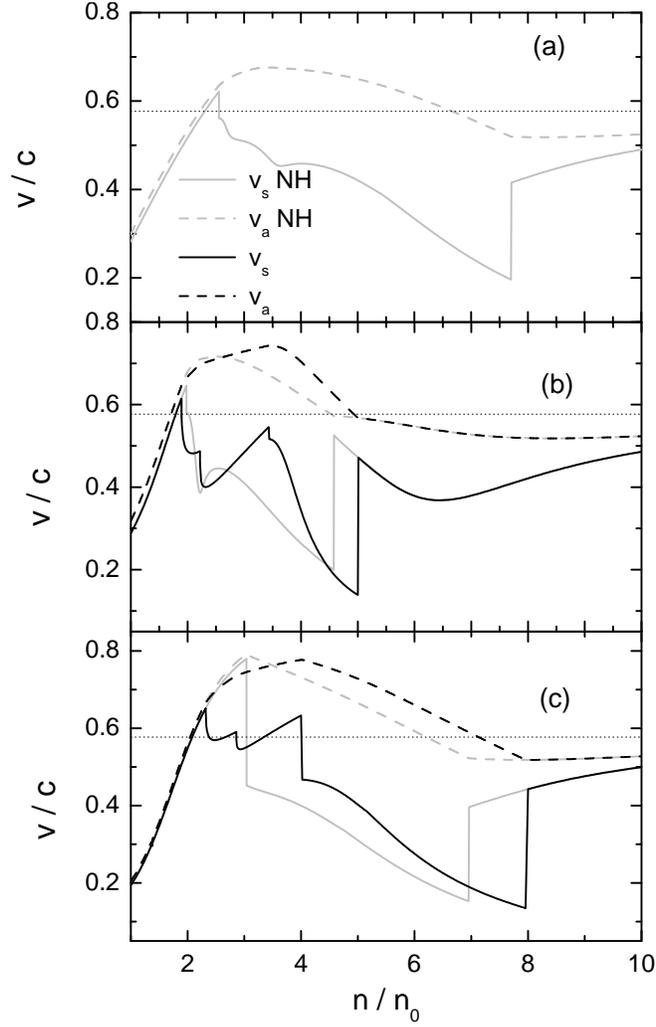}
\caption{\footnotesize The speed of sound as a function of the
baryonic density for the hadronic models GM1e (a), NL3e (b), and
M$\sigma \delta$ (c). The different definitions equilibrium $v_s$
and adiabatic $v_a$ corresponding to the cases with or without
hyperons (NH) have been distinguished according to the line
convention shown. The conformal limit is represented by a
horizontal dotted line.}
\end{figure}

\newpage
\begin{figure}
\includegraphics[width=0.8\textwidth]{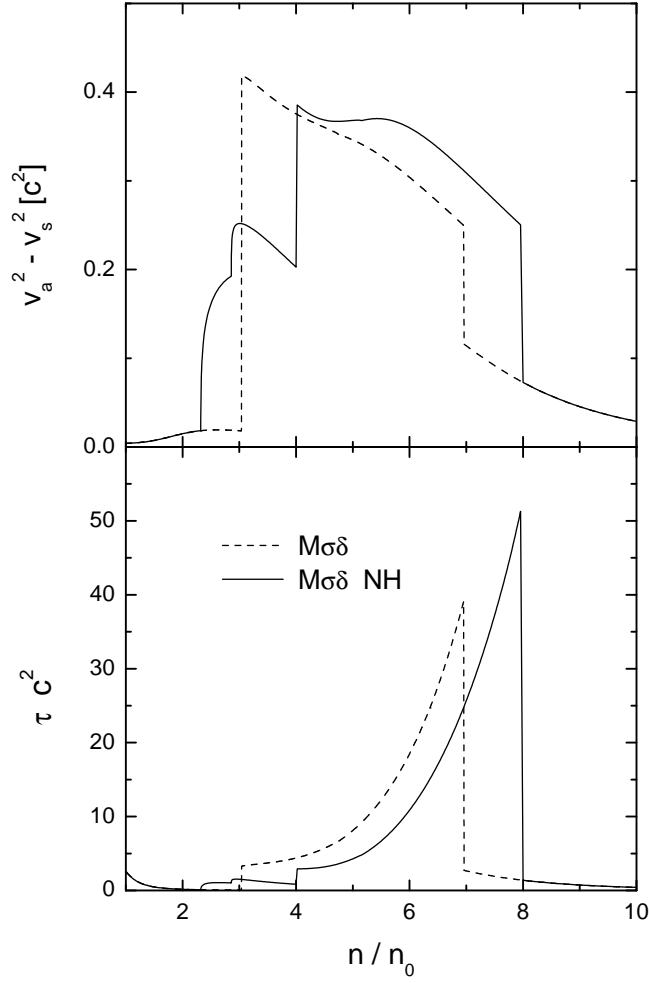}
\caption{\footnotesize The combinations of the different
definitions of the speed of sound $v_a^2-v_s^2$ (upper panel) and
$\tau=1/v_s^2-1/v_a^2$ (lower panel) as functions of the baryonic
density. The cases with or without hyperons (NH) have been
distinguished according to the line convention shown. }
\end{figure}

\newpage
\begin{figure}
\includegraphics[width=0.8\textwidth]{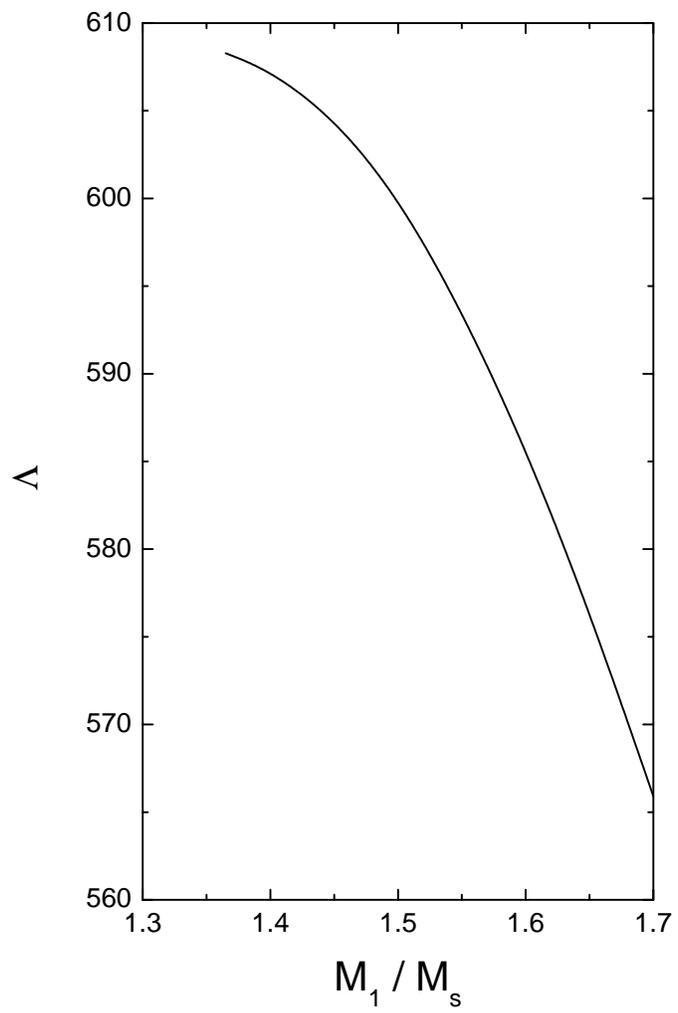}
\caption{\footnotesize The combined tidal deformability as a
function of the mass of the heavier component of a binary system
for the M$\sigma \delta$  model. }
\end{figure}

\end{document}